# Magnetically induced transmission/absorption in chiral liquid crystals

A.H. Gevorgyan

Far Eastern Federal University, 10 Ajax Bay, Russky Island, Vladivostok 690922, Russia; e-mail: agevorgyan@ysu.am

We report the discovery, theoretically, of new effects, namely, the effects of magnetically induced transparency/absorption. The effects are observed in a magnetically active helically structured periodical medium. Changing the external magnetic field and absorption, one can tune the frequency and the linewidth of the transparency/absorption band. We have shown that both effects can be simultaneously observed for two eigenmodes of a cholesteric liquid crystal (CLC) layer. Both effects are nonreciprocal and tunable. They can be observed in the case of the absence of dispersion of the dielectric and the magnetic tensor components, as well as in the case of dispersion dependences of optical parameters of the CLC layer. Comparison of these effects show that while the effect of magnetically induced transparency is observed at wide intervals of variation of the parameters of the CLC layer, the effect of magnetically induced absorption can observe only in very small intervals of these parameters.

Keywords: cholesteric liquid crystals, magnetically induced transparency, magnetically induced absorption, diffraction, eigen modes

## 1. Introduction

Electromagnetically induced transparency (EIT) arises from quantum destructive interference and leads to emergence of a narrow-band transparency window for the light propagating in an absorbing medium [1]. The effect of electromagnetically induced absorption (EIA) which is the counter phenomenon of invoking constructive interference between multiple interaction pathways to enhance/induce absorption, has been proposed and observed [2]. In EIA as well as in EIT, modification of the amplitude spectral response accompanies dispersion tuning. This, in its turn, has some applications such as: slow light [3,4]; non-reciprocal light storage and control [5-7] and microwave photonic delay and switching [8-10]. In [11] it was suggested and theoretically demonstrated the effect of magnetically induced transparency (MIT) for magnetized plasma by Hur et. al. Later, it was confirmed experimentally in [12]. Since that preliminary work, a great deal of theoretical and experimental efforts showed an equivalent transparency effect in gyro-active media in photonic nano structures and other designs [13]. Recently the term MIT has been used to describe the optical effects of high refractive index dielectric nanostructures with magnetic resonances, too [14,15]. In [16] we reported about a MIT effect in opaque chiral liquid crystals (CLCs), using an external magnetic field. In [17] a realization of a new hybrid magneto-plasmonic thin film structure was demonstrated which resembles the classical optical analog of EIA.

As in the case of EIA, we should also expect the existence of the effect of magnetically induced absorption (MIA) in a magnetically active helically structured periodical medium. This work is devoted to such effects. A detailed comparative analysis of the effects of MIT and MIA is carried out, too.

## 2. Models and methodology. Results

We consider light propagation through a planar CLC, that has magneto-optical activity in an external magnetic field directed along the axis of the CLC helix (Fig. 1). We assume that the tensors of dielectric permittivity and magnetic permeability have the form:

$$\hat{\varepsilon}(z) = \varepsilon_m \begin{pmatrix} 1+\delta\cos 2az & \pm\delta\sin 2az \pm ig/\varepsilon_m & 0 \\ \pm\delta\sin 2az \mp ig/\varepsilon_m & 1-\delta\cos 2az & 0 \\ 0 & 0 & 1-\delta \end{pmatrix}, \hat{\mu}(z) = \hat{I}, \quad (1)$$

where $\varepsilon_m = (\varepsilon_1 + \varepsilon_2)/2, \delta = \frac{(\varepsilon_1 - \varepsilon_2)}{(\varepsilon_1 + \varepsilon_2)}$, $\varepsilon_{1,2}$ are the principal values of the local dielectric permittivity tensor in the presence of an external magnetic field; $g$ is the parameter of magneto-optical activity; $a = 2\pi / p$, $p$ is the pitch of the helix in the presence of an external magnetic field. We assume also that light propagates along the CLC optical axis. First, we consider the case when the real and imaginary parts of the principal values of the local dielectric permittivity tensor as well as the magnetooptical activity parameter g are constant and do not depend on frequency.

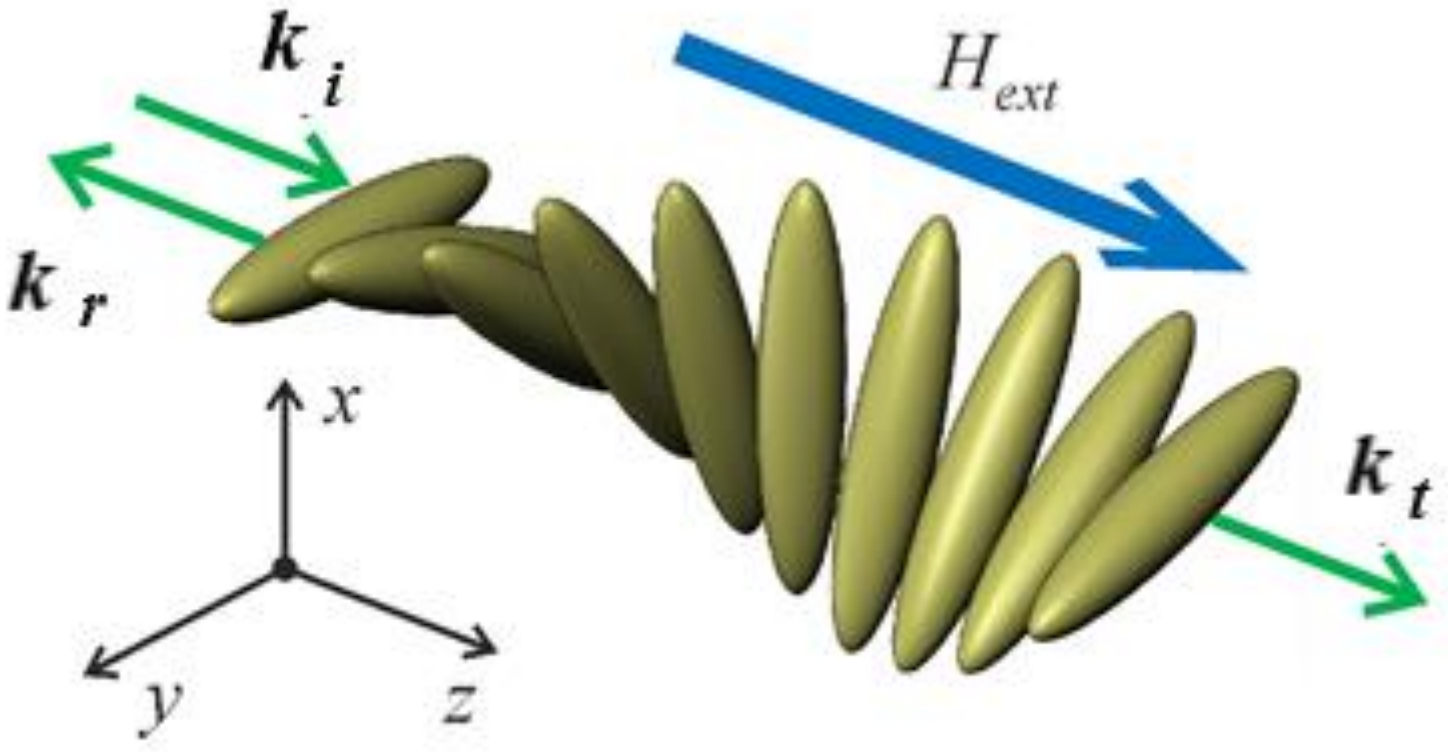

Fig. 1. The geometry of the problem. The large ellipsoids represent the anisotropic molecules, which are rotating continuously, forming a helicoidal structure along the z-axis.

The exact analytical solution of Maxwell's equations for CLCs in an external magnetic field in the rotating frame (the *x*' and *y*' axes of which rotate together with the structure; besides, the *x*' axis is oriented along the local optical axis everywhere, the *y*' axis is perpendicular to the *x*' axis and *z*' axis directed along medium's helix axis [17,18]) when light propagates along the CLC axis, is known (see, in particular [19]). According to [19], the dispersion equation has the form:

$$k_{mz}^4 + a_1 k_{mz}^2 + a_2 k_{mz} + a_3 = 0, \quad (2)$$

where $a_1 = -2\left(\frac{\omega^2}{c^2}\varepsilon_m + a^2\right)$, $a_2 = -4\frac{\omega^2}{c^2}ag$, $a_3 = -2\frac{\omega^2}{c^2}a^2\varepsilon_m + \frac{\omega^4}{c^4}\varepsilon_m^2(1-\delta^2) - \frac{\omega^4}{c^4}g^2 + a^4$, $\omega$ is the angular frequency and $c$ is the speed of light in a vacuum.

Fig. 2 shows the dependences of (a) $\mathrm{Re}k_{mz}$ and (b) $\mathrm{Im}k_{mz}$ on the wavelength $\lambda$ for the following two cases: g=0.15 (the external magnetic field is present, and it is parallel to the light propagation direction; solid lines) and g=0 (the external magnetic field is absent; dashed lines). As show our calculations and as it is seen on Fig. 1, in the absence of an external magnetic field, the curves $\mathrm{Re}k_{mz}$ and $\mathrm{Im}k_{mz}$ are symmetric about the axis $k_{mz} = 0$, which is due to the reciprocity of the system in the absence of an external magnetic field (the curves for $\mathrm{Re}k_{mz}$, in the case g=0, we have not presented, in order not to entangle the figure too much; they have no special features and pass between the curves 1 and 2 and 3 and 4, consequently, as for $\mathrm{Im}k_{mz}$). Further, as it can be seen from Fig. 2b, for the $\mathrm{Im}k_{mz}$ for the two forward eigenmodes (with $\mathrm{Im}k_{mz} > 0$ and $\mathrm{Re}k_{mz} > 0$), some interesting features are observed; namely, in the region near the wavelength λ = 20 nm, the first of them passes through a peak while the second one through a pit with a minimum $\mathrm{Im}k_{mz} = 0$. In addition, for this wavelength (we denote it by $\lambda_t$) we have $\mathrm{Re}k_{1z} = \mathrm{Re}k_{2z}$ (Fig.2a). Note that the two backward modes do not have any specific features except, of course, the well-known ones [7].

Using expressions for $k_{mz}$, we can solve the problem of reflection, transmission, absorption, and localization of light in the case of a magnetoactive CLC layer of a finite thickness. We assume that the optical axis of the CLC layer is perpendicular to the boundaries of the layer and is directed along the *z*-axis. The CLC layer on its both sides borders with isotropic half-spaces with the same refractive indices equal to $n_s$. The boundary conditions, consisting of the continuity of the tangential components of the electric and magnetic fields, are a system of eight linear equations with eight unknowns.

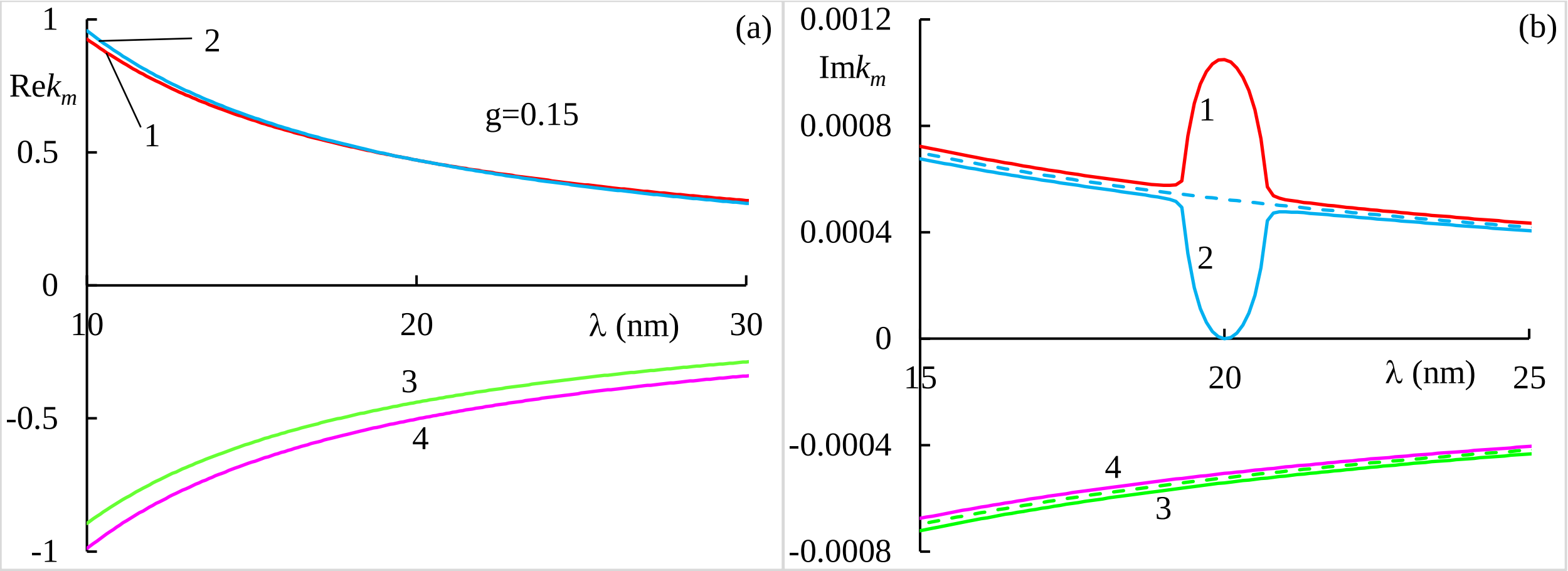

Fig.2. The dependences of (a) Re$k_{mz}$ and (b) Im$k_{mz}$ (*m*=1,2,3,4) for the wavelength $\lambda$ in the following two cases: g=0.15 and g=0. The CLCs parameters are: $\varepsilon_1 = 2.25 + i0$; $\varepsilon_2 = 2.25 + i0.01$; *p*=400 nm.

Thus, solving this boundary-value problem, we can determine the values of the components of the reflected $\vec{E}_r(z)$ and transmitted $\vec{E}_t(z)$ fields, as well as of the field $\vec{E}_{in}(z)$ in the CLC layer itself and, therefore, calculate the reflection $R = {|E_r|^2}/{|E_i|^2}$, transmission $T = {|E_t|^2}/{|E_i|^2}$, and absorption $A = 1 - (R + T)$ coefficients. To minimize the influence of the dielectric boundaries, we consider the case $n_s = \sqrt{\varepsilon_m}$, that is, when the CLC layer is sandwiched between the two half-infinite isotropic spaces with refractive indices equal to the CLC average refractive index. We investigate the specific properties of reflection, transmission, and absorption for the incident light with eigen polarizations (EPs). The EPs are the two polarizations of the incident light that are not changed when light transmits through the system. It follows from the definition of the EPs that they are connected with the polarizations of the excited internal waves (the eigenmodes).

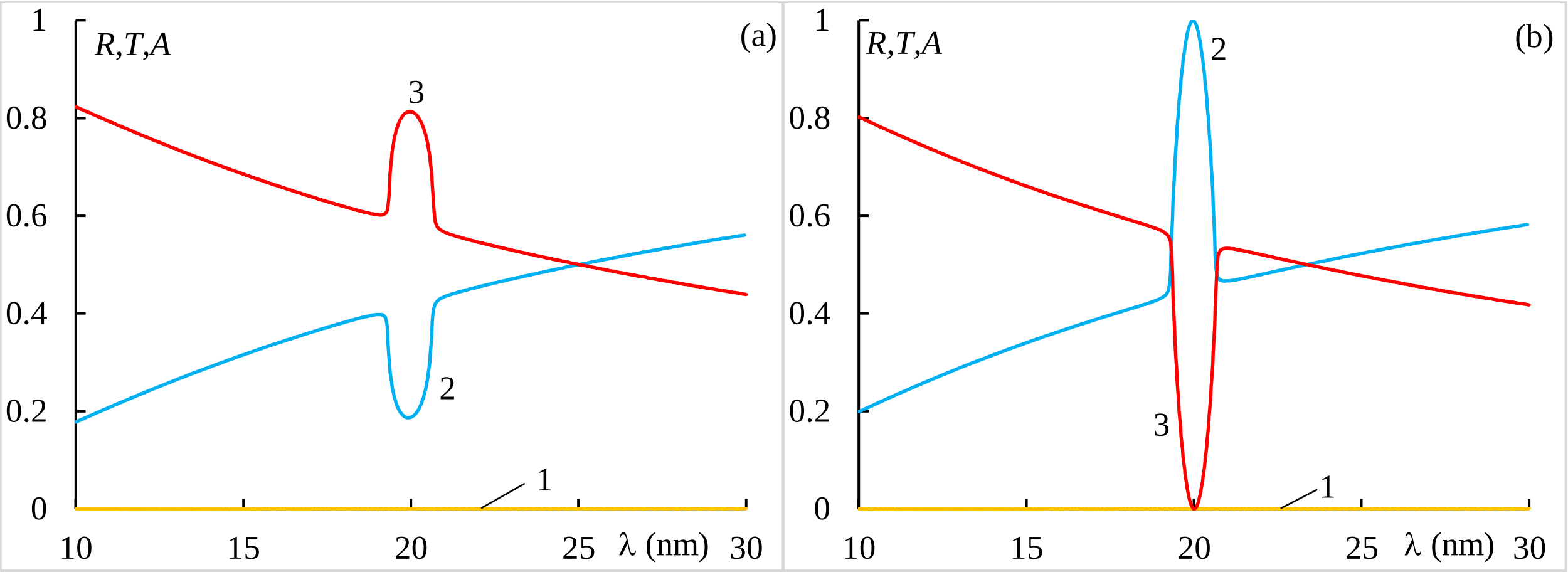

Fig. 3. The spectra of *R* reflection (curve 1), of *T* transmission (curve 2) and of *A* absorption (curve 3) of the first (a) and the second (b) eigenmodes of the CLC layer with the thickness *d*=2*p* and g=0.15. The other parameters are the same as on Fig.2.

Fig.3 shows the reflection, transmission, and absorption spectra for the two EPs of the CLC layer.

From the presented reflection, transmission, and absorption spectra, it follows that:

1) on the wavelength $\lambda_t$, a resonance increasing of the absorption and a resonance decreasing of the transmission take place for the first eigenmode;

2) on the same wavelength $\lambda_t$, for the second eigenmode a transparency window emerges, where $A = 0$ and $T = 1$.

Let us note that these effects emerge only in the presence of the external magnetic field (at $g \neq 0$) and they are observed outside the diffraction reflection region, usually far from it for short wavelengths. These effects are nonreciprocal; they take place for $g > 0$, but do not take place for $g < 0$. That is why we name the first effect MIA in the first case and MIT in the second case. Then these effects do not take place for isotropic absorption, that is, in the case $\text{Im}\varepsilon_1 = \text{Im}\varepsilon_2$.

Now we investigate the influence of the CLC layer with various parameters on the discovered effects and compare these influences for these two effects.

First, we investigate the influence of magnetooptical parameter g on the spectra of absorption for these two EPs.

Fig.4 a,b show the spectra of absorption for the two eigenmodes with MIA and MIT for different values of the parameter g. As follows from these spectra, we can tune the wavelength $\lambda_t$, by changing g, that is, these effects are tunable. For completeness, on Fig. 4 c,d, we show the evolution of the absorption spectra of these two eigenmodes vs a change in the parameter g.

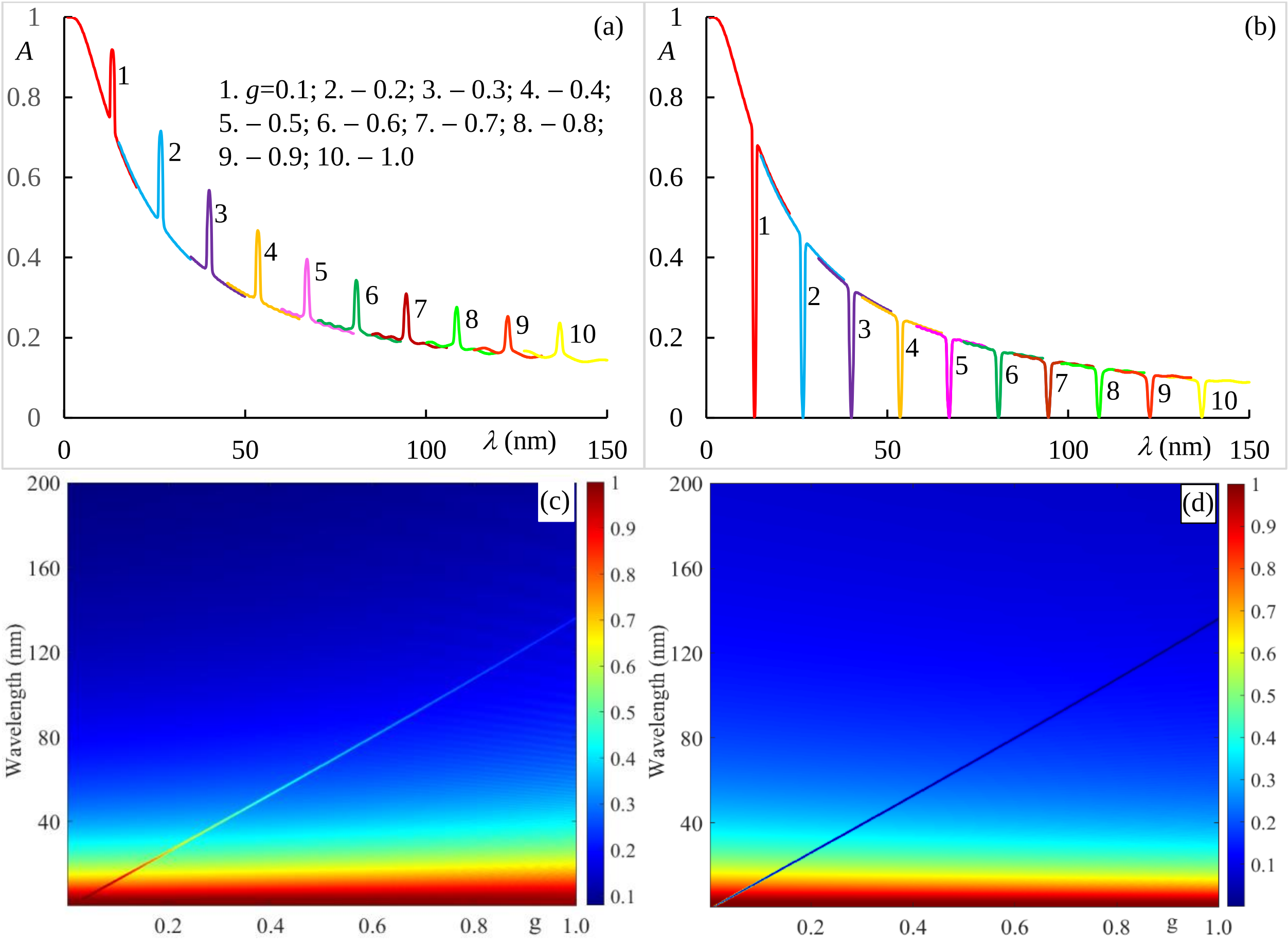

Fig.4. The spectra of absorption for the two eigenmodes with MIA and MIT for different values of the parameter g (a,b) and the evolution of the absorption spectra of these two eigenmodes vs a change in the parameter g (c,d). The parameters are the same as on Fig.3.

Now we investigate the influence of the CLC layer thickness on these effects. Fig.5 a,b show the spectra of absorption for the two eigenmodes with MIA and MIT for different thicknesses of the CLC layer. As follows from these spectra, with an increase in the thickness of the CLC layer, the height of the MIA resonance lines decreases, and, at the same time, the line half-width of these resonances increases, and for large values of the CLC layer thickness this effect practically does not manifest itself. And vice versa, with an increase in the CLC layer thickness, the height of the MIT resonance lines increases, and, at the same time, the line half-width of these resonances decreases. Again, for completeness, on Fig. 5 c,d, we show the evolution of the absorption spectra of these two eigenmodes vs a change in the CLC layer thickness.

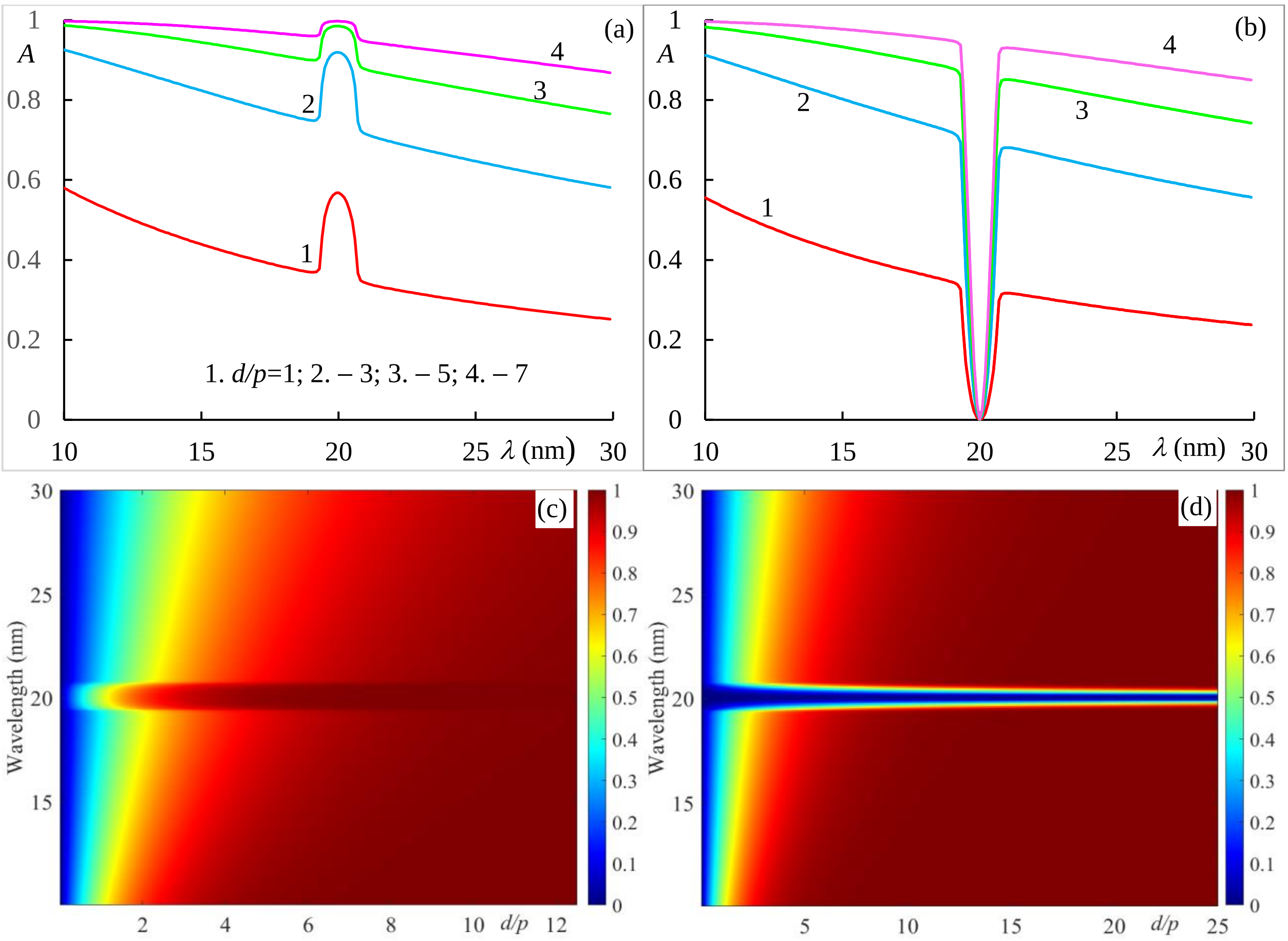

Fig.5. The spectra of absorption for the two eigenmodes with MIA and MIT for different thicknesses of the CLC layer (a,b) and the evolution of the absorption spectra of these two eigenmodes vs a change in the CLC layer thickness (c,d). g=0.15. The other parameters are the same as on Fig.3.

Let us now investigate the influence of ImΔ (at $|\mathrm{Im}\Delta| = |\mathrm{Im}\varepsilon_m|$, that is, when the imaginary part exists only for one principal value of the local dielectric tensor and is absent for the other; $\Delta = \frac{\varepsilon_1 - \varepsilon_2}{2}$) on these effects. Fig.6 a,b show the spectra of absorption for the two eigenmodes with MIA and MIT for various values of ImΔ. As follows from these spectra, with an increase in ImΔ, the height of the MIA resonance lines decreases, and, at the same time, the line half-width of these resonances increases, and for large values of the ImΔ does not manifest itself. While with an increase in the ImΔ, the height of the MIT resonance lines increases, and, at the same time, the line half-width of these resonances increases, too. For completeness, we show the evolution of the absorption spectra of these two eigenmodes with a change in the ImΔ on Fig. 6 c,d.

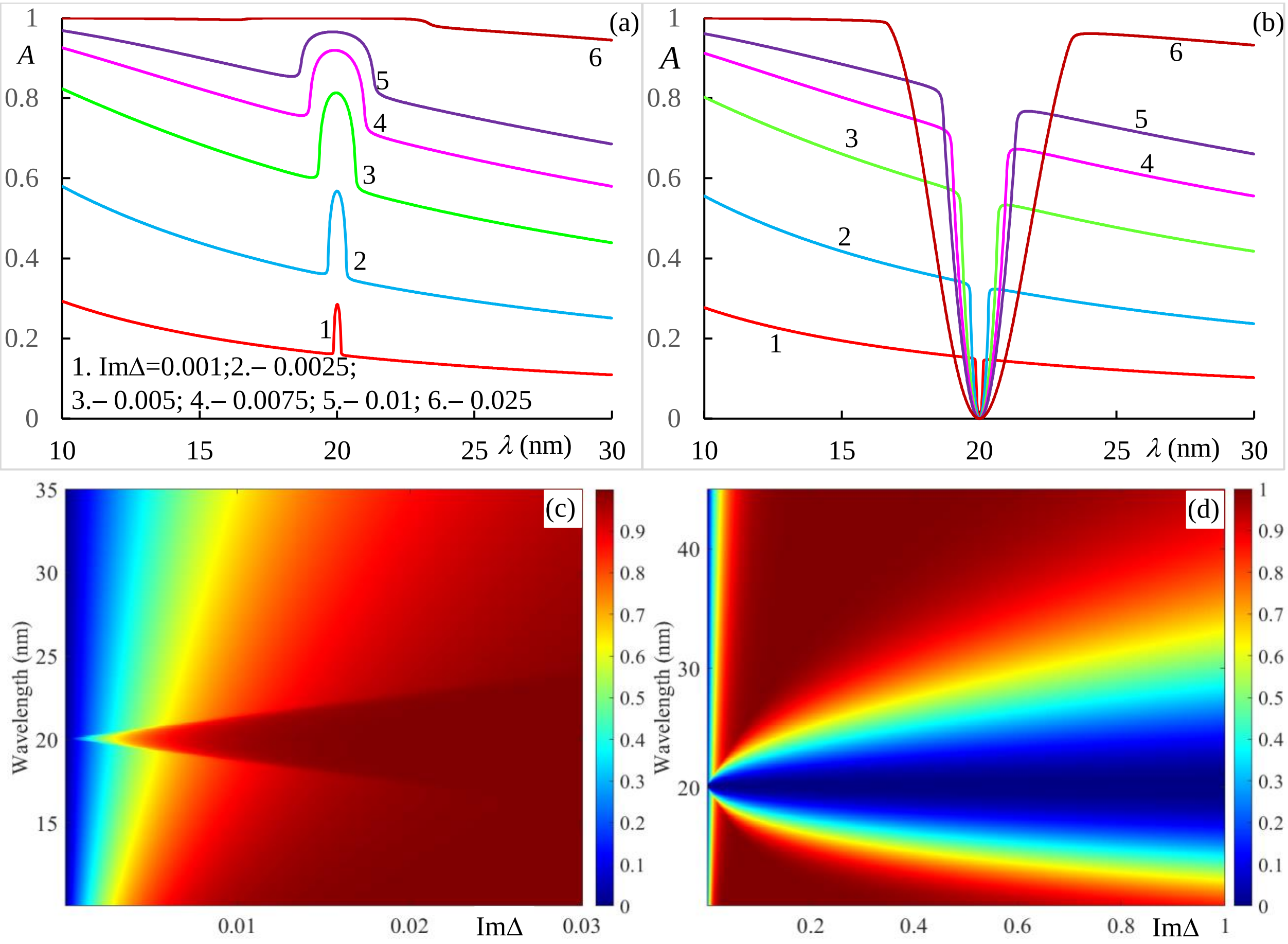

Fig.6. The spectra of absorption for the two eigenmodes with MIA and MIT for different ImΔ (at $|\mathrm{Im}\Delta| = |\mathrm{Im}\varepsilon_m|$) (a,b) and the evolution of the absorption spectra of these two eigenmodes vs a change in the ImΔ (c,d). g=0.15. The other parameters are the same as on Fig.3.

Now we investigate the influence of ReΔ on these effects. Fig.7 a,b show the spectra of absorption for the two eigenmodes with MIA and MIT for different values of ReΔ. As above, with an increase in ReΔ, the height of the MIA resonance lines decreases, and, at the same time, the line half-width of these resonances increases, and for large values of the ReΔ it practically does not manifest itself. While with an increase in the ReΔ, the height of the MIT resonance line decreases, and, at the same time, the line half-width of these resonances increases, too. For completeness, on Fig. 7 c,d we show the evolution of the absorption spectra of these two eigenmodes vs a change in the ReΔ.

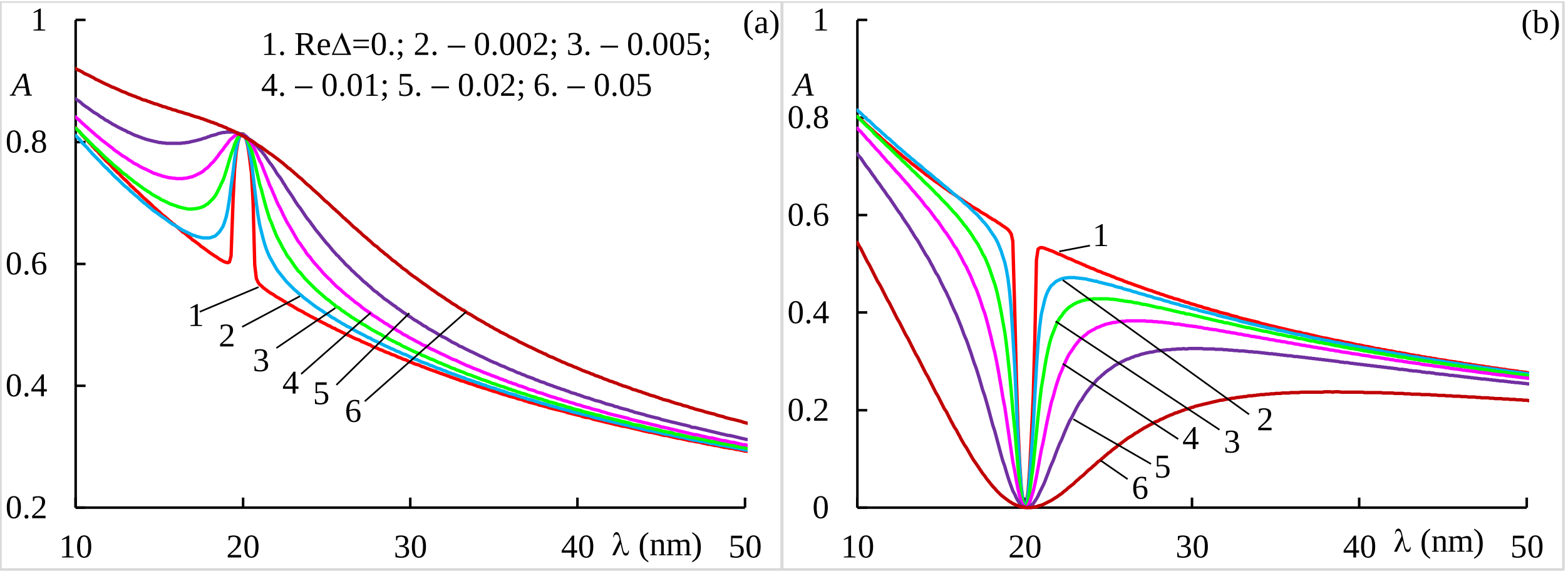

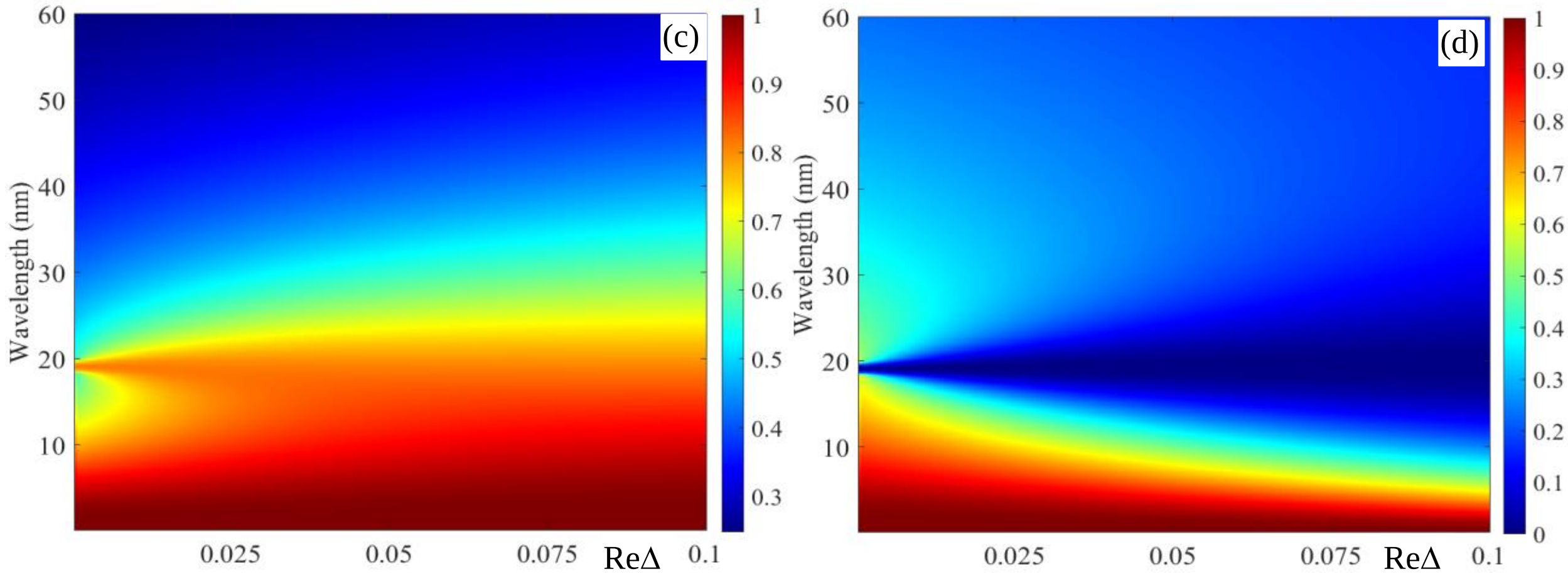

Fig.7. The spectra of absorption for the two eigenmodes with MIA and MIT for various ReΔ (a,b) and the evolution of the absorption spectra of these two eigenmodes vs a change in the ReΔ (c,d). g=0.15. The other parameters are the same as on Fig.3.

Up to now we have supposed that the real and imaginary parts of the dielectric and the magnetic tensor components were constant and did not depend on the frequency. Now we pass to a more general case, considering the dispersion dependences of optical parameters of the CLC layer. We assume that the dielectric permittivity and the magnetic permeability tensors have forms (1) and the principal values, $\varepsilon_{1,2}$, of the local dielectric permittivity tensor have a Lorentzian dependence on the frequency, i. e. we assume that $\varepsilon_{1,2}$ in (1) have the forms

$$\varepsilon_{1,2}(\omega) = \varepsilon_{1,2}^{0} + \frac{f_{1,2}}{\omega_{01,2}^{2} - \omega^{2} - i\gamma_{1,2}\omega}, \quad (3)$$

while, as above, $\mu_{1,2}$, are constant and do not depend on the frequency. Here $\gamma_{1,2}$ are the broadenings of resonance absorption lines or simply the damping factors, $f_{1,2}$ are quantities proportional to oscillator strengths, $\omega_{01,2}$ are the resonance frequencies, and $\varepsilon_{1,2}^{0}$ are the parts of the dielectric permittivities, not depending on the frequency and *i* is the imaginary unit. Fig. 8 shows the spectra of (a) the transmission $T$, and (b) the absorption $A$ for the two eigenmodes with MIA and MIT for g=0.15 (solid lines 1 and 2 correspondingly) and for g=0 (in this case the transmission and absorption spectra for these two modes coincide: dashed lines). As follows from the presented results, against the background of a broadband absorption/transmission line in the absence of an external magnetic field, some ultra-narrow-band absorption lines emerge for the MIA mode as well as some ultra-narrow-band lines of induced transmission for the MIT mode. So, MIT/MIA can tailor the amplitude and phase response of an absorption resonance to create large dispersion, which has been exploited for applications in slow- and fast-light applications and in quantum information science. Since MIA enhances (induces) absorption of an already existing absorption (transmission) profile, it leads to a significant reduction in the output signal power. Note that in the case of MIA, the nature of the narrowness of the absorption line in a resonantly amplifying medium can compensate for the loss of the signal, while creating a large dispersion at the MIA frequency. In its turn, narrow absorption features within an MIT resonance may find application in light pulse shaping and simultaneous filtering of multiple signals.

As our simulations show, the same results could be expected for other dispersion laws of the dielectric constant.

We now turn to a discussion of the question of how an absorbing medium can provide full transmission of light along a narrow spectral line without any absorption (or how a line of one hundred percent suppression of absorption for transmitted light in an absorbing medium can emerge) or how a super-strong absorption line can exist in a medium with relatively weak absorption.

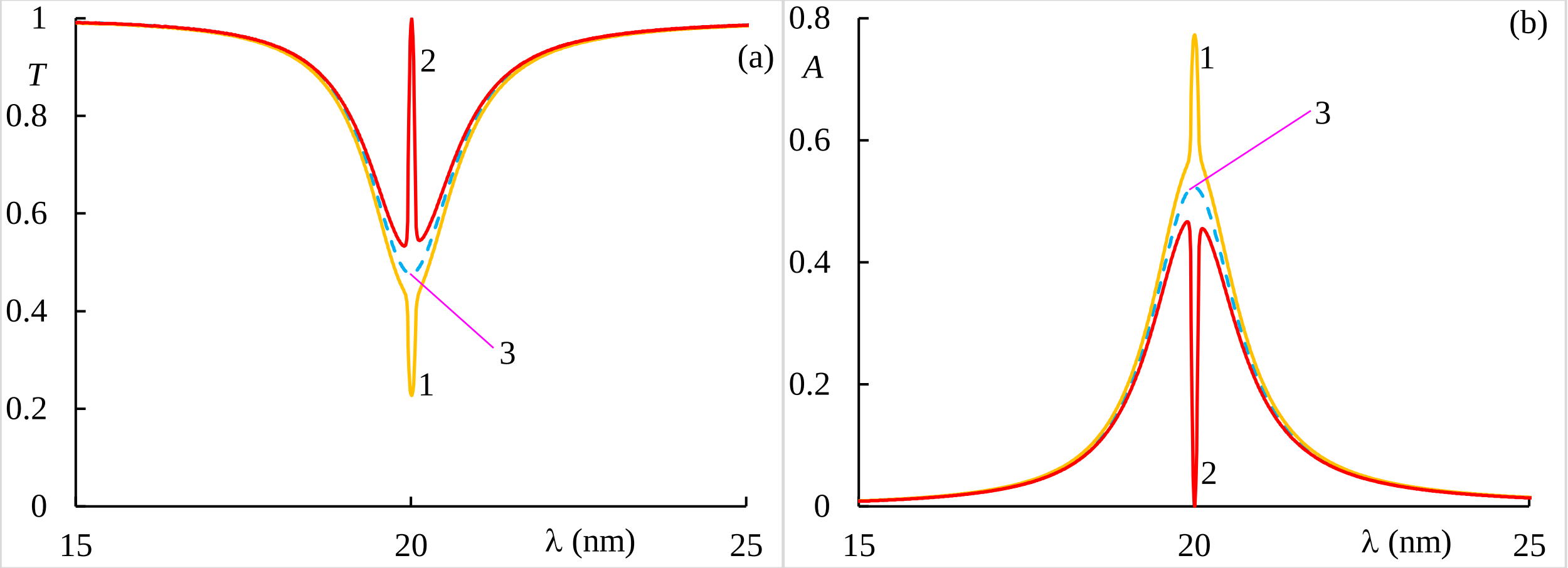

Fig. 8. (a) The transmission $T$ and (b) the absorption $A$ spectra for the two eigenmodes with MIA and MIT for g=0.15 (solid lines 1 and 2 correspondingly) and for g=0 (in this case the transmission and absorption spectra for these two modes coincide: dashed lines). The rest of the parameters are: g=0.15, and $d$=20$p$, $f_1 = 0$, $f_2 = 5 \cdot 10^{29}\ \mathrm{s}^{-2}$, $\gamma_2 = 6 \cdot 10^{15}\ \mathrm{s}^{-1}$, $\omega_{02} = 2\pi c/\lambda_{02}$, $\lambda_{02} = 20$ nm.

To answer these questions, we pass to investigate the specific properties of the total wave exiting in the CLC layer for the two modes with MIA and MIT. Presenting the total electric field corresponding to the medium on the left side of the CLC layer, in the layer itself and in the medium on the right side of the layer (we assume that the CLC layer is located between two isotropic half-spaces $z$ = 0 and $z$ = $d$) in the form

$$\vec{E} = \begin{cases} \vec{E}_i(z) + \vec{E}_r(z), z < 0, \\ \hat{R}^{-1}(az)\vec{\mathcal{E}}(z), 0 < z < d, \\ \vec{E}_t(z), z > d, \end{cases} \tag{4}$$

we also investigate the evolution of the spectra of the amplitude and polarization characteristics of this field as a function of $z/p$ for the two above said eigen modes of the CLC layer.

The total standing wave exiting in the periodically inhomogeneous medium is modulated. Fig.9 a,b show the dependences of the ellipticity $e_{\mathrm{total}}$ of the total wave exiting in the medium for the incident light of the wavelength $\lambda_t$ vs the parameter $z/p$, where $p$ is the pitch of the helix and $z$ is the coordinate of the axis directed along the axis of the medium for the eigen mode with MIA and for the eigen mode with MIT. As follows from these dependences, in both cases the total waves for this wavelength have practically linear polarizations with $e_{\mathrm{total}} \approx 0$; moreover, $e_{\mathrm{total}}$ makes modulated oscillations with an increasing amplitude of modulation around the axis $e_{\mathrm{total}} = 0$ in the first case (a) and near the same axis in the second case (b). The study of the dependence of the azimuth $\varphi$ on $z$ shows that it has a linear character and, moreover, for a change in $z$ equal to the pitch of the helix, there is a change in the azimuth $\varphi$ equal to 2π and for both eigenmodes. For completeness, on Fig. 9 c,d, we show the evolution of the dependences of the ellipticity $e_{\mathrm{total}}$ of the total wave exiting in the medium on the parameter $z/p$ vs a change in the wavelength λ.Thus, it can be stated that this situation is like the one which occurs in the anisotropic absorption near the photonic band gap (PBG) edges, and which ensures the existence of the effects of anomalously strong absorption and complete suppression of absorption [20]. So, for the eigenmode with MIA, the total wave has linear polarization and is wrapped up to the CLC helix and, everywhere in the CLC layer, is directed along the direction of the coincidence with the direction of the maximum absorption, while for the eigenmode with MIT it is directed along the direction of the coincidence with the direction of the minimum absorption. It also follows that the one hundred percent transmission is possible only under the condition Im$\Delta = \pm\mathrm{Im}\varepsilon_m$, and also explains why these effects do not take place in the case of isotropic absorption.

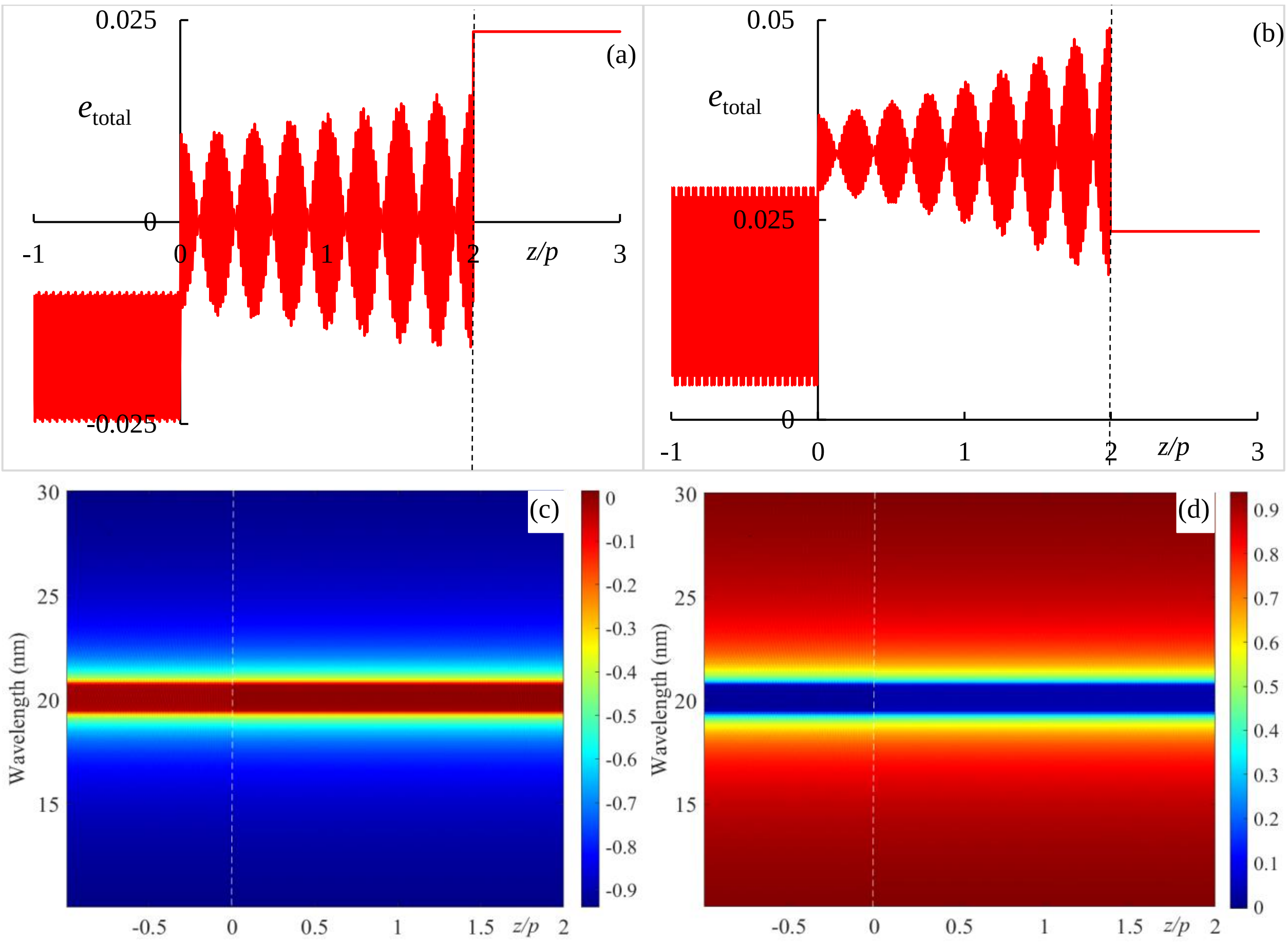

Fig.9. The dependence of the ellipticity of the total wave exiting in the medium on the parameter *z/p*, where *p* is the pitch of the helix and *z* is the coordinate of the axis directed along the axis of the medium for the eigen mode with MIA (a) and one with MIT (b). The parameters are the same as on Fig.3.

Fig. 10 shows the density plots of the spectra of $|\mathbf{E}(z)|^2$ as functions of $z/p$ for the two eigen modes with MIA (a) and MIT (b).

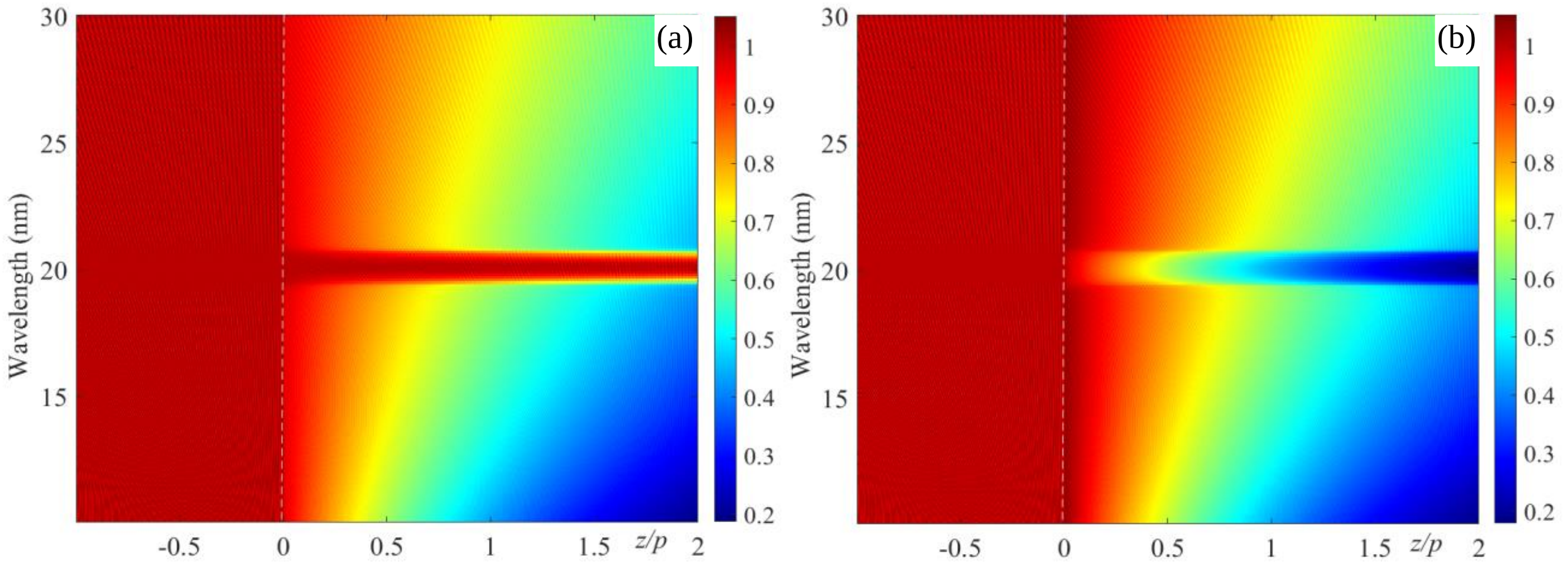

Fig. 10. The density plots of the spectra of $|\mathbf{E}(z)|^2$ as functions of $z/p$ for the two eigen modes with MIA (a) and MIT (b). The parameters are the same as on Fig.3.

## 3. Conclusions

In conclusion, we investigated the specific peculiarities of effects MIT and MIA in the CLC layer with magneto-optical activity. First, we investigated the dependences of the real and imaginary parts of wave vectors of the eigen modes on the wavelength and showed that for g>0 there exists some wavelength $\lambda_t$ for which the first of them passes through a peak while the second one goes through a pit with a minimum $\mathrm{Im}k_{mz} = 0$. In addition, for this wavelength $\lambda_t$ and for these two eigen modes, we have $\mathrm{Re}k_{1z} = \mathrm{Re}k_{2z}$. Then we investigated the specific peculiarities of $R$ reflection, $T$ transmission, and $A$ absorption spectra near the wavelength $\lambda_t$. Here $R$ has not any specific peculiarities and $R \approx 0$, because $\lambda_t$ is far from PBG. Then, for the wavelength $\lambda_t$, some resonance increase of the absorption and resonance decrease of the transmission take place for the first eigenmode and, for the second eigenmode, there emerges a transparency window, where $A = 0$ and $T = 1$ for this wavelength $\lambda_t$. So, in the CLC layer with magnetooptical activity there can emerge two effects simultaneously, namely, MIA for the one eigen mode and MIT for the second one. We investigated the CLC parameter changes influence on these effects, namely, that for the change of magnetooptical parameter g, as well as that for the CLC layer thickness $d$ change and, finally, that for the real and imaginary parts changes. Comparison of these effects for these changes show that while the effect of MIT is observed in wide intervals of variation of these parameters of the CLC layer, the effect MIA can be observed only in very small intervals of variation of these parameters. The investigation of the influence of the dispersion dependences of optical parameters of the CLC layer show, that MIT/MIA can tailor the amplitude and the phase response of an absorption resonance to create large dispersion, which has been exploited for applications in slow- and fast-light applications, as well as in quantum information science. Narrow absorption features within an MIT resonance may find application in light pulse shaping and simultaneous filtering of multiple signals. Moreover, the capability for creating a transparency window and for tuning the linewidth of such resonance and transparency band frequency is important for such applications as tunable bandwidth filters, etc. In conclusion, we discussed the question of how an absorbing medium can provide full transmission of light along a narrow spectral line without any absorption, or how a super-strong absorption line can exist in a medium with relatively weak absorption. We showed, that for these two modes the total standing wave exiting in the periodically inhomogeneous medium is linearly polarized and, moreover, for the eigenmode with MIA, the total wave is wrapped up about the CLC helix and directed along the direction of the coincidence with the direction of the maximum absorption everywhere in the CLC layer, while for the eigenmode with MIT it is directed along the direction of the coincidence with the direction of the minimum absorption.